\documentclass[journal, final]{IEEEtran}

\usepackage{graphicx}
\usepackage{subfigure}
\usepackage{lipsum}
\usepackage{amstext}
\usepackage{amsmath}
\usepackage{amssymb}
\usepackage[dvipsnames]{xcolor}
\usepackage[hyphens]{url}

\usepackage{algorithm,algorithmic}

\begin{document}

\title{A Federated Filtering Framework for Internet of Medical Things}

\author{
	\IEEEauthorblockN{
		Sunny Sanyal\IEEEauthorrefmark{1},
		Dapeng Wu\IEEEauthorrefmark{1},
		and Boubakr Nour\IEEEauthorrefmark{2}
	}\\
	\IEEEauthorblockA{\IEEEauthorrefmark{1}Chongqing University of Posts and Telecommunications, China}\\
	\IEEEauthorblockA{\IEEEauthorrefmark{2}School of Computer Science, Beijing Institute of Technology, Beijing, China}\\
	Email: sanyal.sunny111@ieee.org, wudp@cqupt.edu.cn, n.boubakr@bit.edu.cn

\thanks{© 2019 IEEE.  Personal use of this material is permitted.  Permission from IEEE must be obtained for all other uses, in any current or future media, including reprinting/republishing this material for advertising or promotional purposes, creating new collective works, for resale or redistribution to servers or lists, or reuse of any copyrighted component of this work in other works.}

}

\markboth{IEEE International Conference on Communications (IEEE ICC 2019), Accepted}%
{Cui \MakeLowercase{\textit{et al.}}: Performance Analysis of Massive SM MIMO in HSR}

\maketitle

\begin{abstract}
	Based on the dominant paradigm, all the wearable IoT devices used in the healthcare sector also known as the internet of medical things (IoMT) are resource constrained in power and computational capabilities. The IoMT devices are continuously pushing their readings to the remote cloud servers for real-time data analytics, that causes faster drainage of the device battery. Moreover, other demerits of continuous centralizing of data include exposed privacy and high latency. This paper presents a novel Federated Filtering Framework for IoMT devices which is based on the prediction of data at the central fog server using shared models provided by the local IoMT devices. The fog server performs model averaging to predict the aggregated data matrix and also computes filter parameters for local IoMT devices. Two significant theoretical contributions of this paper are the global tolerable perturbation error (${To{l_F}}$) and the local filtering parameter ($\delta$); where the former controls the decision-making accuracy due to eigenvalue perturbation and the later balances the tradeoff between the communication overhead and perturbation error of the aggregated data matrix (predicted matrix) at the fog server. Experimental evaluation based on real healthcare data demonstrates that the proposed scheme saves upto 95\% of the communication cost while maintaining reasonable data privacy and low latency.

\end{abstract}

\IEEEpeerreviewmaketitle

\section{Introduction}
World Health Organization (WHO) recently reports~\cite{who2017} a global health workforce shortage of 12.9 million during the coming decade. This expected shortage accompanied by various other factors have inspired a slow but steady paradigm shift from conventional healthcare to the smart healthcare~\cite{islam2015internet, hammami2018proactive}. The smart healthcare enables patients with round the clock monitoring and feedback and is also expected to automate critical operations inside ICU~\cite{gholami2018ai}.  Internet of Things (IoT) is widely accepted~\cite{baker2017internet} as a crucial driver to the connected healthcare paradigm. Allied Market Research predicts~\cite{who2018} a global market capital for IoT healthcare to reach 136.8 billion US dollar by 2021, moreover today we already have 3.7 million connected internet of medical things (IoMT) devices.

A typical wearable IoMT device consists of a tiny battery which in most cases is nonchargeable~\cite{higginbotham2018internet}, and this leads to disposal of the equipment once it is out of charge. A significant cause of speedier disposition of IoMT devices is due to the dominant cloud computing paradigm~\cite{armbrust2010view} of pushing all the collected data to the distant cloud servers for analytics and decision making. This phenomenon incurs a significant loss of power due to high communication overhead. Moreover, it also exposes the aggregated sensitive medical data to the security risks. This paper considers the problem of high power loss, exposure medical data privacy and high latency in cloud based healthcare analytics. It is an interesting problem as it has social implications also governments~\cite{who2012} and industries (Cisco~\cite{cisco2014}, Microsoft~\cite{microsoft2014}) are investing a lot of money and resources to develop a future healthcare infrastructure.

\begin{figure}[b]
	\centering
	\includegraphics[width=0.8\linewidth]{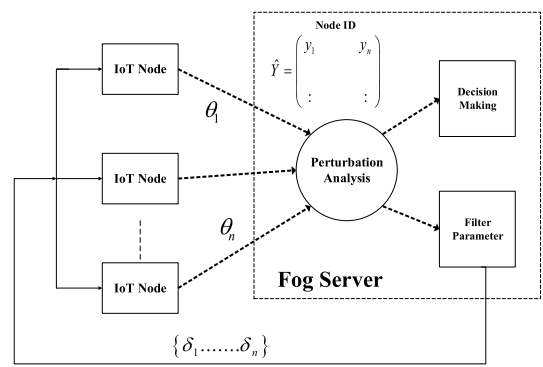}
	\caption{Federated Filtering Framework.}
	\label{fig:1}
\end{figure}

This paper presents an algorithmic framework namely Federated Filtering Framework (FFF) (Fig.~\ref{fig:1}) for IoMT supported by theoretical analysis. The proposed framework presents an alternate solution to the issues of energy efficiency, latency and privacy for resource-constrained IoMT devices. In brief, each IoMT device computes a local model of the data and shares this model with the fog server. The fog server's role is threefold. First, it predicts a data matrix (aggregated data matrix) using aggregated model average (Section V); second, it computes and delivers filter parameters for all the IoMT devices and finally performs decision making using the aggregated data matrix. To control the eigenvalue perturbation of the data matrix that may compromise the decision accuracy this paper derives a theoretical relation between the local filtering parameter with the global tolerable eigenvalue perturbation using Matrix Perturbation Theory (MPT).

Overall, the contributions of the paper are as follows: (i) a theoretical relationship between local time series filtering and perturbation error of aggregated data matrix (ii) the implementation of federated decision making framework using filters, (iii) a lightweight fully unsupervised local subroutine (algorithm 1), (iv) the filter model averaging (algorithm 2) that preserves the privacy and demands few updates, (v) a practical framework for IoMT data aggregation.

The article is organised in the following fashion. Section II discusses the related work. Section III presents the system model. Section IV presents the theoretical analysis. Section V presents the kernel of the paper which is Federated Filtering Framework. Section IV shows the experimental evaluation, and finally, the article concludes by highlighting the significant contributions and future work.

\section{Related Work}
This section compares the proposed framework with three closely related genres of research that includes IoT in healthcare, prediction based IoT systems and federated learning approaches in networks.

\subsection{IoT in Healthcare}
The dominant paradigm for IoT based healthcare analytic systems~\cite{baker2017internet} can be categorized as cloud computing-based health monitoring and mobile computing based health monitoring. Both the scenarios mentioned above very frequently push data to the server (cloud server/mobile device) for decision making. This paper is firmly against the continuous transmission of data and presents a prediction based data aggregation scheme with error bounds to ensure the fidelity of the decision making. Some recent use cases of IoT based healthcare analytics such as~\cite{ammae2018unobtrusive, mora2017iot} also advocates centralized decision making, however, both of them lacks a theoretical formulation to ensure decision-making accuracy.

\subsection{Prediction based IoT systems}
The literature~\cite{dias2016importance} reports several prediction based approaches for reducing the communication overhead in sensor networks. The prediction~\cite{dias2016survey} based approaches are categorized into single prediction approaches and dual prediction approaches. In the case of single prediction approaches the system performs prediction in only one location whereas in the case of  dual prediction approaches the system performs prediction at a local node along with the central server. Some notable prediction schemes applicable for both the categories mentioned above are adaptive filtering scheme~\cite{fathy2018adaptive}, Autoregressive filter, Autoregressive Integrated Moving Average filter (ARIMA)~\cite{dias2016survey}, Kalman filtering and machine learning techniques~\cite{wu2016data}. Although some of the prior approaches can provide better accuracy for the model generation at the IoT device however given the severe computational constraints of the IoMT devices these approaches are not practical for local processing. Moreover, none of the earlier approaches shows any relationship between local and global processing using theoretical upper bounds.

\subsection{Federated Learning in Networks}
The effectiveness of federated averaging algorithm for distributed training proposed by Mcmahan \textit{et al.}~\cite{mcmahan2017communication} provides strong motivation to develop a federated filtering framework for IoMT devices. Moreover, there are also other notable distributed optimization approaches~\cite{yang2013trading, nour2017distributed} that improves communication efficiency. All the distributed and federated approaches in the literature are highly complex to run in a tiny IoMT device furthermore, they aim to perform decision making at the user device. The proposed Federated Filtering Framework on the other hand proposes a very lightweight subroutine for the local IoMT device and also aims to perform decision making at the server using local shared model.

\section{System Model}
The system model considers a massive IoMT scenario where $n$ number of IoMT devices are cumulatively working towards sensing a particular phenomenon. All the IoMT devices are connected to the fog server(s) using Wi-Fi links. Each IoMT device $\left\{ {{N_1}, \dots, {N_n}} \right\} \in {N_i}$ generates a time series data stream. This paper assumes a centrally aggregated matrix $Y$ also known as global matrix (real) of size $m \times n$ where each column (${Y_i}$ ) represents a particular IoMT device, and each row has a sensor reading of every 30 seconds. This generation of a global matrix $Y$ requires continuous transmission of data to the fog server. However, this paper doesn't advocate a continuous push and therefore proposes a prediction based framework. The fog server generates an aggregated data matrix ($\hat Y$); i.e. a predicted data matrix with perturbation and as earlier ${\hat Y_i}$ represents a column vector of the data matrix. The perturbation in the global data matrix is due to errors caused by filtering and predicton. The formation of aggregated data matrix is discussed in Section V. The fog server's role is threefold. Firstly it estimates/predicts the perturbed data matrix ($\hat Y$), and secondly it computes and delivers filter parameter (${\delta _i}$) for all the IoMT devices, and finally, it performs decision making using the perturbed data matrix. Table 1 shows some important notations.

\begin{table}[t]
	\begin{center}
		\caption{The description of main symbols.}
		\begin{tabular}{|p{40pt}|l|p{60pt}|}
			\hline
			Symbol&
			Description \\
			\hline
			${N_i}$ &
			
			${i^{th}}$ IoMT device  \\
			\hline
			${Y}$ &
			
			Global matrix\\
			\hline
			${{Y_i}(t)}$ &
			
			The ${i^{th}}$ column of the global matrix\\
			\hline
			$\hat  \cdot$ &
			
			Perturbed version of the original symbol\\
			\hline
			${\delta _i}$ &
			
			The ${i^{th}}$ filter parameter \\
			\hline
			
			${\theta _i}$  &
			
			Prediction model of ${i^{th}}$ IoMT data\\
			\hline
			$e$ &
			
			Mean square error function\\
			\hline
			$\alpha$  &
			
			Learning rate/step size \\
			\hline
			$\lambda$ &
			
			Eigen value of a matrix\\
			\hline
			$\Delta$ &
			
			The perturbation error \\
			\hline
		\end{tabular}
		\label{tab1}
	\end{center}
\end{table}

In the beginning, all the IoMT nodes train the prediction model by running several instances of Least Mean Square (LMS) filter (section IV A). Both the local IoMT device and the fog server uses the same prediction scheme. The local IoMT device runs a local processing subroutine as described in Algorithm 1 and the fog server runs Algorithm 2.

\begin{algorithm}[t]
	\caption{Local Processing Protocol}
	\label{alg:1}
	\begin{algorithmic}[1]
		\REQUIRE current ${\theta _i}(t)$, ${\delta _i}(t)$, ${Y_i}(t)$, ${\alpha _i}(t)$
		\ENSURE $\theta _i^*(t)$
		\FOR {(true)}
		\STATE t= current time
		\STATE compute: ${W_i}(t) = {Y_i}(t) - {\hat Y_i}(t)$
		\IF {$\left| {{W_i}(t)} \right| > {\delta _i}$}
		\STATE $[\theta _i^*(t)]: = LMS({Y_i}(t),{\alpha _i}(t))$  $\leftarrow$ Eq. 3
		\STATE ${N_i}$ sends (i, $\theta _i^*(t)$, ${Y_i}(t)$) to fog server
		\STATE Set ${W_i}(t)$ $\leftarrow$ 0
		\STATE Set  ${\theta _i}(t)$ $\leftarrow$ $\theta _i^*(t)$
		\ENDIF
		\ENDFOR
	\end{algorithmic}
\end{algorithm}

\section{Theoretical Analysis}

\subsection{Adaptive Filtering at IoMT Devices}
Adaptive filters are typically implemented for signals with non-stationary statistics and where no prior information is available. A typical adaptive filter is depicted in Fig. 2. Among various adaptive filters~\cite{haykin2008adaptive} this paper selects Least Mean Square (LMS) filter~\cite{haykin2003least} for local processing inside the IoMT node, since it has a very low computational overhead.

\begin{figure}[b]
	\centering
	\includegraphics[width=0.7\linewidth]{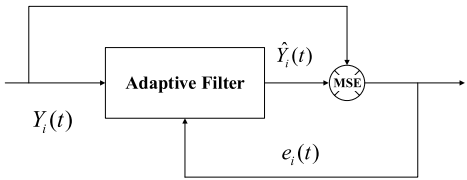}
	\caption{Typical Adaptive Filter}
	\label{Fig.2._}
\end{figure}

Let for an IoMT device ${N_i}$ at time t the predicted IoMT sensor vector ${\hat Y_i}(t)$ be a linear approximation of the real sensor vector ${Y_i}(t)$. The LMS adaptive filter embedded inside the IoMT devices aims to minimise error the function $e(t)$, which is the least mean square approximation between the predicted sensor vector and the real sensor vector.
\begin{equation}
{e_i}(t) = \frac{1}{2}\sum\limits_{i = 1}^n {{{\left( {{{\hat Y}_i}(t) - {Y_i}(t)} \right)}^2}}
\end{equation}

The relationship between the predicted sensor vector  ${\hat Y_i}(t)$ (output of LMS filter) and the real sensor vector ${Y_i}(t)$ is as follows.
\begin{equation}
{\hat Y_i}(t) = \theta _i^T{Y_i}(t)
\end{equation}

The LMS filter relies on the stochastic gradient descent (SGD) optimisation, this approach takes iterative steps (${\alpha _i}$) towards the steepest decrease of the error function ${e_i}(t)$. Eq. 3 shows the LMS update rule also known as Widrow-Hoff learning rule.
\begin{equation}
{\theta _i}(t) = {\theta _i}(t - 1) + {\alpha _i}(t) \cdot {e_i}(t) \cdot {Y_i}(t)
\end{equation}

Based on the empirical observation~\cite{haykin2008adaptive} to ensure convergency the step size ${\alpha _i}(t)$ should satisfy the following.

\begin{equation}
0 \le {\alpha _i}(t) \le \frac{1}{{{P_Y}}}
\end{equation}

where ${P_Y} = \frac{1}{M}\sum\limits_{j = 1}^M {{{\left| {{Y_i}(j)} \right|}^2}}$, and M is the number of iterations taken for training the LMS filter.

\subsection{Perturbation Analysis at Fog Server}
The filter parameters play a key role in balancing the tradeoff between the desirable loss of decision accuracy (by allowing perturbation to $\hat Y$) and low communication overhead. This paper uses the matrix perturbation theory~\cite{stewart1990matrix} to bound the perturbation error ($\Delta$) of the perturbed data matrix which in turn affects the decision accuracy. The fog server generates a perturbed data matrix $\hat Y = Y + W$, where $W$ is the perturbation/filtering error and column elements of $W$,  ${W_i} \in \left[ { - {\delta _i},{\delta _i}} \right]$. Let the ${\lambda _i}$ and ${\hat \lambda _i}$ denote the eigenvalues of the real covariance matrix $A = \frac{1}{m}{Y^T}Y$ and the perturbed covariance matrix $\hat A = \frac{1}{m}{\hat Y^T}\hat Y$ respectively.

The norm of the perturbation error matrix $\Delta  = A - \hat A$ can be formulated using the property of triangle inequality~\cite{huang2006distributed} is depicted as follows.
\begin{equation}
\begin{split}
\left\| \Delta  \right\| = \left\| {{Y^T}W + {W^T}Y + {W^T}W} \right\| \\
\le \left\| {{Y^T}W} \right\| + \left\| {{W^T}Y} \right\| + \left\| {{W^T}W} \right\|
\end{split}
\end{equation}
The goal here is to determine an upper bound for the expectation of RHS in the above inequality.

This paper assumes that all the column vectors of $W$ are independent and all the column elements are i.i.d random variables with zero mean ($\mu  = 0$) and variance $\sigma _i^2 \approx \sigma _i^2({\delta _i})$ along with fourth moment as $\mu _i^4 = \mu _i^4({\delta _i})$.

Using Jensen inequality $E(x) \le \sqrt {E({x^2})}$.
\begin{equation}
\begin{split}
E({\left\| \Delta  \right\|_F}) \le 2E\left( {{{\left\| {{Y^T}W} \right\|}_F}} \right) + E\left( {{{\left\| {{W^T}W} \right\|}_F}} \right) \\
\le 2\sqrt {E\left( {\left\| {{Y^T}W} \right\|_F^2} \right)}  + \sqrt {E\left( {\left\| {{W^T}W} \right\|_F^2} \right)}
\end{split}
\end{equation}

Based on Mirsky's theorem~\cite{stewart1990matrix}.
\begin{equation}
E\left( {\sqrt {\frac{1}{n}\sum\limits_{i = 1}^n {{{\left( {{{\hat \lambda }_i} - {\lambda _i}} \right)}^2}} } } \right) \le E\left( {\frac{{{{\left\| \Delta  \right\|}_F}}}{n}} \right) \le To{l_F}
\end{equation}

\begin{equation}
\begin{split}
E({\left\| \Delta  \right\|_F}) \le 2\sqrt {\frac{1}{{{m^2}n}}Tr\left( {{Y^T}Y} \right) \cdot \sum\limits_{i = 1}^n {\sigma _i^2} }  \\
+ \sqrt {\left( {\frac{1}{m} + \frac{1}{n}} \right) \cdot \sum\limits_{i = 1}^n {\sigma _i^4} }
\end{split}
\end{equation}

\begin{equation}
E({\left\| \Delta  \right\|_F}) \le To{l_F}
\end{equation}

The Eq. 9 presents an upper bound ($To{l_F}$) on the perturbation error caused due to local filtering at IoMT devices and  estimation of perturbed data matrix using outdated shared model.
\begin{equation}
\begin{split}
To{l_F} = 2\sqrt {\frac{1}{{{m^2}n}}Tr\left( {{Y^T}Y} \right) \cdot \sum\limits_{i = 1}^n {\sigma _i^2} }  \\
+ \sqrt {\left( {\frac{1}{m} +  \frac{1}{n}} \right) \cdot \sum\limits_{i = 1}^n {\sigma _i^4} }
\end{split}
\end{equation}
Similar upper bounds can also be derived using spectral norm ${\left\| . \right\|_2}$, moreover this paper selects Frobenious norm ${\left\| . \right\|_F}$ for no particular reason.

\subsection{Uniform filter parameter selection}
This paper assumes an independent and uniform distribution of IoT filter parameter within the interval $\left[ { - {\delta _i},{\delta _i}} \right]$. Moreover we also assume a homogeneous filter parameter allocation among all the IoMT devices, therefore ${\delta _i} = \delta$ and ${\sigma _i} = \frac{{{\delta ^2}}}{3}$. Solving Eq. 10 for $\delta$.

\begin{equation}
\begin{split}
\delta  = \frac{{\sqrt {\frac{{3Tr({Y^T}Y)}}{m} + 3 \cdot To{l_F} \cdot \sqrt {nm + {m^2}} }  - \sqrt {\frac{{3 \cdot Tr({Y^T}Y)}}{m}} }}{{\sqrt {m + n} }}
\end{split}
\end{equation}
The Eq. 11 provides a relationship between local filtering and the global perturbation error, that plays a crucial role in balancing the tradeoff between local filtering at IoMT devices and the global eigen perturbation error.

\section{Federated Filtering Framework}

Federated Filtering Framework (FFF) since the system is based on a loose federation of the participating devices (IoMT devices) those are coordinated by the central fog server. The FFF consists of two crucial protocols first, the local data processing protocol and the second is global data processing and coordination protocol.

\subsection{Local Processing Protocol at IoT Device}
Given the severe resource constraints in computation for IoMT devices, this paper proposes a very lightweight filtering protocol for local processing. The local filtering is based on LMS adaptive filter (section). The IoMT devices computes a local prediction model ${\theta _i}$ (Eq. 10) from the collected data and share this model with the fog server. Now assuming ${\theta _i}$ as the current prediction model and ${\delta _i}$ as the latest filtering parameter for the ${N_i}$ IoMT device. ${N_i}$ at any time t tracks the deviation of predicted sensor vector ${\hat Y_i}(t)$ from real sensor vector ${Y_i}(t)$ using ${W_i}(t) = {Y_i}(t) - {\hat Y_i}(t)$. Whenever $\left| {{W_i}(t)} \right| > {\delta _i}$ the IoMT device updates the prediction model ${\theta _i}(t)$ and resets ${W_i}\left( t \right)$ to zero. The updated prediction model along with a small amount of sample data is shared with the fog server. However the LMS filter incurs negligible computational overhead that enables the IoMT device to run multiple instances of filtering for better accuracy. The above mentioned details for local processing at IoMT devices is summarized in algorithm 1.

\subsection{Federated Processing at Fog Server}
At the beginning of each round the fog servers updates the current prediction models with the newly shared models. The fog server selects a random fraction K of the n participating IoMT devices. This paper selects a random fraction of IoMT devices~\cite{mcmahan2017communication} since the decision accuracy degrades beyond a certain number. The step size ${\alpha _i}(t)$ is kept constant based on the empirical result (section IV). The fog server aggregates the model using Eq. 12.
\begin{equation}
{\eta _i}(t)= \sum\limits_{i = 1}^K {\frac{{{n_k}}}{n}{\theta _i}(t - 1)}
\end{equation}

Thereafter the fog server predicts the perturbed data matrix using the following equation.
\begin{equation}
\hat Y(t) = \eta _i^TY(t)
\end{equation}

The perturbed data matrix $\hat Y(t)$ is used for decision making. The impact of eigen perturbation error on the decision accuracy can be studied in~\cite{huang2006distributed}. The fog server continuously tracks  $E({\left\| \Delta  \right\|_F}) > To{l_F}$ . Once the data matrix perturbation error exceeds the tolerable perturbation error threshold, the fog shares an updated filter parameter and summons all IoMT devices to share their updated prediction model. The above mentioned scheme is summarized as Algorithm 2.
\begin{algorithm}[t]
	\caption{Filter Model Averaging}
	\label{alg:2}
	\begin{algorithmic}[1]
		\FOR {(true)}
		\STATE t~=~current time
		\IF {$E({\left\| \Delta  \right\|_F}) \le To{l_F}$}
		\STATE  ${\eta _i}(t)$ $\leftarrow$ $\sum\limits_{i = 1}^K {\frac{{{n_k}}}{n}{\theta _i}(t - 1)}$
		\STATE $\hat Y(t)$ $\leftarrow$  $\eta _i^TY(t)$  $\leftarrow$ Eq. 2
		\STATE \textbf{Perform} decision making
		\ELSE
		\STATE Fog server shares ${\delta _i}$ with ${N_i}$
		\STATE Fog server receives (i, $\theta _i^*(t)$, ${Y_i}(t)$)
		\ENDIF
		\ENDFOR
	\end{algorithmic}
\end{algorithm}

\textbf{Advantages:} The proposed framework minimises the communication overhead (section VI) by limiting the number of transmissions to the central server. The algorithm 2, i.e. the model averaging makes it practically impossible to extract an individual model from the average model; that ensures privacy to sensitive medical data. Furthermore, the fog server, unlike a cloud server, is located closer to the source, which reduces the latency.

\section{Performance Evaluation}
In this section, we present some experimental results based on real IoT health data. The results include the prediction using the filter model averaging by the fog server, the plot of communication overhead while varying local filtering parameter and the overall scalability of the proposal concerning energy efficiency. The experiments are performed using a publicly available\footnote{http://archive.ics.uci.edu/ml/datasets/MHEALTH+Dataset} real IoT health dataset known as MHEALTH (mobile health) data. The dataset comprises body motion and vital signs recordings for ten volunteers of diverse profile while performing 12 physical activities. For our experiments, we have only considered the chest accelerometer sensor reading, i.e. columns 1-3 and the right lower arm gyro sensor time series data, i.e. column 18-20.

Based on section IV/C we assume a homogeneous filter parameter for all the IoMT devices. We initially distribute the data equally among 50 IoMT devices and compute normalized tolerable perturbation error as shown in Eq. 14.
\begin{equation}
\left\langle {To{l_F}} \right\rangle  = {{To{l_F}} \mathord{\left/
		{\vphantom {{To{l_F}} {\sqrt {\frac{{\sum {\lambda _i^2} }}{n}} }}} \right.
		\kern-\nulldelimiterspace} {\sqrt {\frac{{\sum {\lambda _i^2} }}{n}} }}
\end{equation}
We present the relationship between the normalized tolerable perturbation error and the uniform filter parameter in Fig. 3.  Moreover Fig. 3 depicts a roughly linear relationship between the normalized tolerable perturbation error and the local filter parameter. It is also intuitive since whenever one increases the $\left\langle {To{l_F}} \right\rangle$ , the filter at IoMT devices passes more data.

\begin{figure}[htbp]
	\centering
	\vspace{-0.6cm}
	\includegraphics[width=0.8\linewidth]{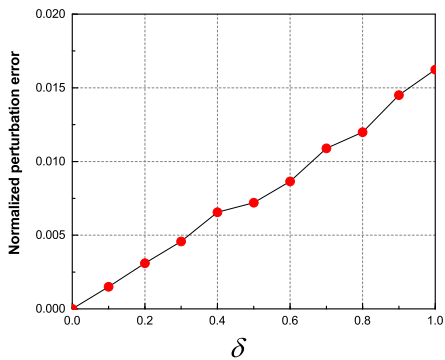}
	\vspace{-0.4cm}
	\caption{Normalized Tolerable Perturbation Error as a function of $\delta$.}
	\label{Fig.3._}
\end{figure}

Next, we present the prediction performance of the filter model averaging scheme (Algorithm 2) by the fog server. Due to space limitations, we offer prediction results of two different IoMT devices (Fig. 4).  As discussed in section both the local and the global filtering uses the same LMS filter. The available sophisticated techniques that provide better accuracy cannot be used at the fog server since those techniques must also be feasible for local processing at IoMT devices. Given the severe resource constraints in power and computation, the sophisticated methods cannot be used by IoMT devices for local processing.

\begin{figure}[h]
	\centering
	\subfigure[Accelerometer sensor readings]{
		\label{fig:T_vs_a} 
		\includegraphics[width=0.45\textwidth]{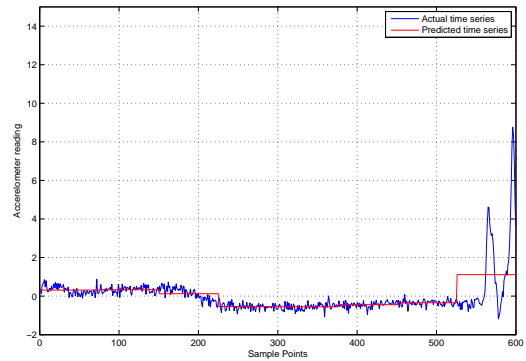}}
	\subfigure[Gyro sensor readings]{
		\label{fig:max_vs_a} 
		\includegraphics[width=0.45\textwidth]{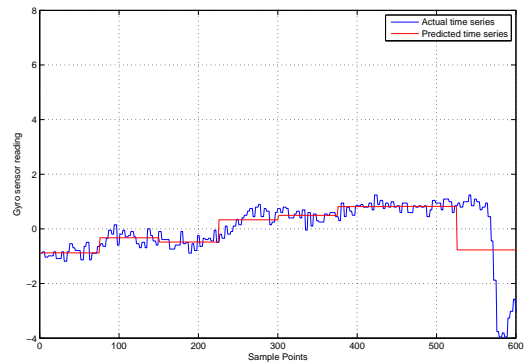}}
	\caption{Prediction performance of Federated Filtering scheme}
	\label{Fig.4} 
\end{figure}

Towards this end, we plot the communication overhead as a function of filter parameter. We observe that in Fig. 5 the communication cost can be massively reduced even with a tolerable perturbation error. We have achieved upto \text{95\%} reduction in transmissions with various tolerable perturbation error. This supports our claim that the proposed  framework can provide a good tradeoff between communication efficiency and eigen perturbation error of data matrix.

Finally, we examine the scalability~\cite{sanyal2017co} of the proposed scheme for small as well as a large number of devices. The energy efficiency ($\eta$ ) of the system with n number of IoMT devices~\cite{sanyal2018improving} can be computed as:	
\begin{equation}
\eta  = \sum\limits_n {\frac{{{d_n}}}{{{E_n} \cdot {r_n} \cdot TTI}}}
\end{equation}
Where ${d_n}$ is the total volume of data to be uploaded, ${E_n}$  is the average energy consumed to deliver a single packet, ${r_n}$ is total number of data packets to be uploaded by all the IoMT devices  and $TTI = 1$ is transmission time interval which is constant to all packets. It is evident from the plot (Fig. 6) that our FFF scheme is highly scalable compared to other recent researches such as AM-DR~\cite{fathy2018adaptive} and well known ARIMA~\cite{dias2016survey}. Based on the plot, the energy efficiency increases with the number of devices. Therefore the proposed framework can also be extended to a massive IoMT scenario.

\begin{figure}[t]
	\centering
	\vspace{-0.7cm}
	\includegraphics[width=0.5\textwidth]{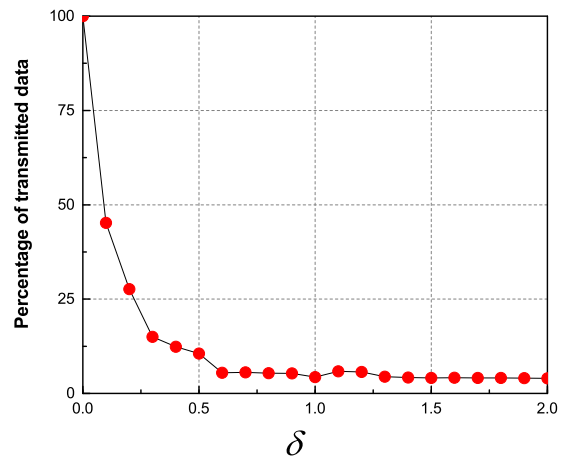}
	\vspace{-0.8cm}
	\caption{Communication overhead as a function of $\delta$.}
	\label{Fig.5._}
\end{figure}

\begin{figure}[t]
	\centering
	\vspace{-0.6cm}
	\includegraphics[width=0.5\textwidth]{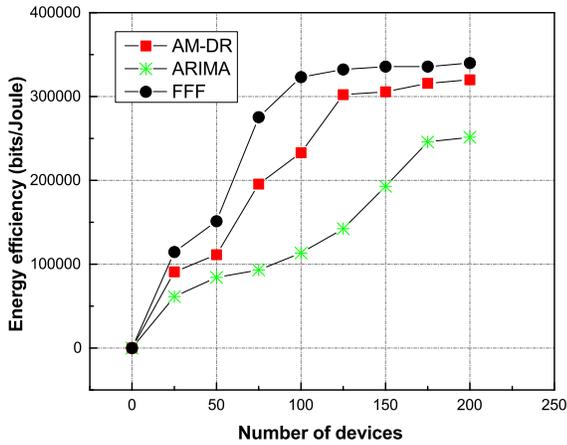}
	\vspace{-0.8cm}
	\caption{Energy efficiency as a function of number of devices.}
	\vspace{-0.4cm}
	\label{Fig.6._}
\end{figure}

\section{Conclusion}
This paper considers open challenges concerning energy efficiency,  privacy and latency for smart healthcare analytics. This paper derives a theoretical upper bound on the eigenvalue perturbation and further formulates a relationship between the local quantization at IoMT devices with the global perturbation error at fog server. Based on the theoretical infrastructure this paper proposes two subroutines first for the local filtering at the IoMT device and the second for the central fog server. The proposed framework cuts down \text{95\%} of the communication overhead. Moreover, the use of perturbed data matrix (predicted data) instead of using real global matrix for decision making ensures better privacy and the low proximity of fog server provides low latency. Future work includes formulating a general relation between decision accuracy and perturbation error and developing an IoMT testbed for verifying the proposed framework.

\section{Acknowledgement}
This work is jointly supported by the National Natural Science Foundation of China under Grant 61871062 and Grant 61771082, also the Program for Innovation Team Building at Institutions of Higher Education in Chongqing under Grant CXTDX201601020. The Authors want to extend their gratitude to Prof. Pingping Zhang of Chongqing University of Posts and Telecommunications, Chongqing, China for her help with the mathematical formulation. Sunny Sanyal is the corresponding author.

\bibliographystyle{IEEEtran}
\bibliography{1570502869Ref}

\begin{thebibliography}{10}
\providecommand{\url}[1]{#1}
\csname url@samestyle\endcsname
\providecommand{\newblock}{\relax}
\providecommand{\bibinfo}[2]{#2}
\providecommand{\BIBentrySTDinterwordspacing}{\spaceskip=0pt\relax}
\providecommand{\BIBentryALTinterwordstretchfactor}{4}
\providecommand{\BIBentryALTinterwordspacing}{\spaceskip=\fontdimen2\font plus
\BIBentryALTinterwordstretchfactor\fontdimen3\font minus
  \fontdimen4\font\relax}
\providecommand{\BIBforeignlanguage}[2]{{%
\expandafter\ifx\csname l@#1\endcsname\relax
\typeout{** WARNING: IEEEtran.bst: No hyphenation pattern has been}%
\typeout{** loaded for the language `#1'. Using the pattern for}%
\typeout{** the default language instead.}%
\else
\language=\csname l@#1\endcsname
\fi
#2}}
\providecommand{\BIBdecl}{\relax}
\BIBdecl

\bibitem{who2017}
``{Global Health Workforce Shortage to Reach 12.9 Million in Coming Decades},''
  Available:
  \url{http://www.who.int/mediacentre/news/releases/2013/health-workforce-shortage/en/},
  {Accessed: 2017-03-27}.

\bibitem{islam2015internet}
S.~R. Islam, D.~Kwak, M.~H. Kabir, M.~Hossain, and K.-S. Kwak, ``{The internet
  of things for health care: a comprehensive survey},'' \emph{IEEE Access},
  vol.~3, pp. 678--708, 2015.

\bibitem{hammami2018proactive}
S.~E. Hammami, H.~Moungla, and H.~Afifi, ``{Proactive Anomaly Detection Model
  for eHealth-Enabled Data in Next Generation Cellular Networks},'' in
  \emph{IEEE International Conference on Communications (ICC)}, 2018, pp. 1--6.

\bibitem{gholami2018ai}
B.~Gholami, W.~M. Haddad, and J.~M. Bailey, ``{AI in the ICU: In the intensive
  care unit, artificial intelligence can keep watch},'' \emph{IEEE Spectrum},
  vol.~55, no.~10, pp. 31--35, 2018.

\bibitem{baker2017internet}
S.~B. Baker, W.~Xiang, and I.~Atkinson, ``{Internet of Things for Smart
  Healthcare: Technologies, Challenges, and Opportunities},'' \emph{IEEE
  Access}, vol.~5, pp. 26\,521--26\,544, 2017.

\bibitem{who2018}
B.~Marr, ``{Why The Internet Of Medical Things (IoMT) Will Start To Transform
  Healthcare In 2018},'' Available:
  \url{https://www.forbes.com/sites/bernardmarr/2018/01/25/why-the-internet-of-medical-things-iomt-will-start-to-transform-healthcare-in-2018/}.

\bibitem{higginbotham2018internet}
S.~Higginbotham, ``{The internet of trash [Internet of Everything]},''
  \emph{IEEE Spectrum}, vol.~55, no.~6, pp. 17--17, 2018.

\bibitem{armbrust2010view}
M.~Armbrust, A.~Fox, R.~Griffith, A.~D. Joseph, R.~Katz, A.~Konwinski, G.~Lee,
  D.~Patterson, A.~Rabkin, I.~Stoica \emph{et~al.}, ``{A view of cloud
  computing},'' \emph{Communications of the ACM}, vol.~53, no.~4, pp. 50--58,
  2010.

\bibitem{who2012}
``{WHO-ITU National e-Health Strategy Toolkit},'' Available:
  \url{http://www.itu.int/ITU-D/cyb/events/2012/eHealth}.

\bibitem{cisco2014}
``{Cisco Services for Connected Health: Intelligent Network, Smart Medicine},''
  Available:
  \url{http://www.cisco.com/web/strategy/healthcare/connectedhealth/index.html}.

\bibitem{microsoft2014}
``{Health Solutions From Microsoft},'' Available:
  \url{http://www.microsoft.com/windowsembedded/en-us/healthcare.aspx}.

\bibitem{ammae2018unobtrusive}
O.~Ammae, J.~Korpela, and T.~Maekawa, ``{Unobtrusive detection of body
  movements during sleep using wi-fi received signal strength with model
  adaptation technique},'' \emph{Future Generation Computer Systems}, 2018.

\bibitem{mora2017iot}
H.~Mora, D.~Gil, R.~M. Terol, J.~Azor{\'\i}n, and J.~Szymanski, ``{An IoT-Based
  Computational Framework for Healthcare Monitoring in Mobile Environments},''
  \emph{Sensors}, vol.~17, no.~10, p. 2302, 2017.

\bibitem{dias2016importance}
G.~M. Dias, B.~Bellalta, and S.~Oechsner, ``{On the importance and feasibility
  of forecasting data in sensors},'' \emph{arXiv preprint}, 2016.

\bibitem{dias2016survey}
------, ``{A survey about prediction-based data reduction in wireless sensor
  networks},'' \emph{ACM Computing Surveys}, vol.~49, no.~3, p.~58, 2016.

\bibitem{fathy2018adaptive}
Y.~Fathy, P.~Barnaghi, and R.~Tafazolli, ``{An Adaptive Method for Data
  Reduction in the Internet of Things},'' in \emph{IEEE World Forum on Internet
  of Things}, 2018.

\bibitem{wu2016data}
M.~Wu, L.~Tan, and N.~Xiong, ``{Data prediction, compression, and recovery in
  clustered wireless sensor networks for environmental monitoring
  applications},'' \emph{Information Sciences}, vol. 329, pp. 800--818, 2016.

\bibitem{mcmahan2017communication}
B.~McMahan, E.~Moore, D.~Ramage, S.~Hampson, and B.~A. y~Arcas,
  ``{Communication-Efficient Learning of Deep Networks from Decentralized
  Data},'' in \emph{Artificial Intelligence and Statistics}, 2017, pp.
  1273--1282.

\bibitem{yang2013trading}
T.~Yang, ``{Trading computation for communication: Distributed stochastic dual
  coordinate ascent},'' in \emph{Advances in Neural Information Processing
  Systems}, 2013, pp. 629--637.

\bibitem{nour2017distributed}
B.~Nour, K.~Sharif, F.~Li, and H.~Moungla, ``{A Distributed ICN-Based IoT
  Network Architecture: An Ambient Assisted Living Application Case Study},''
  in \emph{IEEE Global Communications Conference}, 2017, pp. 1--6.

\bibitem{haykin2008adaptive}
S.~S. Haykin, \emph{{Adaptive filter theory}}.\hskip 1em plus 0.5em minus
  0.4em\relax Pearson Education India, 2008.

\bibitem{haykin2003least}
S.~Haykin and B.~Widrow, \emph{{Least-mean-square adaptive filters}}.\hskip 1em
  plus 0.5em minus 0.4em\relax John Wiley \& Sons, 2003, vol.~31.

\bibitem{stewart1990matrix}
G.~W. Stewart, ``{Matrix perturbation theory},'' 1990.

\bibitem{huang2006distributed}
L.~Huang, X.~Nguyen, M.~Garofalakis, M.~Jordan, A.~Joseph, and N.~Taft,
  ``{Distributed PCA and network anomaly detection},'' in \emph{In Proceedings
  of NIPS}, vol. 2006, 2006.

\bibitem{sanyal2017co}
S.~Sanyal, D.~Wu, J.~Yan, and X.~Li, ``{Co-relative mobility based IoT data
  uploading using D2D communication},'' in \emph{EAI International Conference
  on Mobile Multimedia Communications}, 2017, pp. 170--175.

\bibitem{sanyal2018improving}
S.~Sanyal and P.~Zhang, ``{Improving Quality of Data: IoT Data Aggregation
  Using Device to Device Communications},'' \emph{IEEE Access}, vol.~6, pp.
  67\,830--67\,840, 2018.

\end{thebibliography}

\end{document}